\documentclass[11pt,a4paper]{article}
\usepackage{jheppub}

\usepackage{multirow}
\usepackage{mathrsfs}
\usepackage{amssymb}
\usepackage{verbatim}
\usepackage{rotating} 
\usepackage{bbm}
\usepackage{array}
\usepackage{float}
\usepackage{color}
\usepackage{graphicx}
\def\ra{\rightarrow}

\newcommand{\bea}{\begin{eqnarray}}
\newcommand{\eea}{\end{eqnarray}}
\newcommand{\be}{\begin{equation}}
\newcommand{\ee}{\end{equation}}

\newcommand{\mbTh}{\mathbf{\Theta}}
\newcommand{\mbOm}{\mathbf{\Omega}}

\newcommand{\lhood}{\mathcal{L}}
\newcommand{\ev}{\mathcal{Z}}

\def\DC{{\sc Double Chooz}}
\def\DB{{\sc Daya Bay}}
\def\RENO{{\sc RENO}}
\def\T2K{{\sc T2K}}
\def\MINOS{{\sc MINOS}}
\def\MN{{\sc MultiNest}}

\newcommand{\figref}[1]{Fig.~\ref{#1}}
\newcommand{\tabref}[1]{Tab.~\ref{#1}}
\newcommand{\secref}[1]{Sec.~\ref{#1}}

\newcommand{\ie}{\emph{i.e.}}

\newcommand{\df}{{\rm d}}
\newcommand{\qu}[1]{``#1''}
\newcommand{\eref}[1]{Eq.~(\ref{#1})}
\def\lbrac{\left\lbrace}
\def\rbrac{\right\rbrace}

\newcommand{\refcite}[1]{Ref.~\cite{#1}}

\title{Bayesian evidence for non-zero $\theta_{13}$ and CP-violation in neutrino oscillations}
\author[a]{Johannes Bergstr\"om}
\affiliation[a]{Department of Theoretical Physics, School of Engineering Sciences,\\
KTH Royal Institute of Technology -- AlbaNova University Center,\\
Roslagstullsbacken 21, 106 91 Stockholm, Sweden}
\emailAdd{johbergs@kth.se}
\abstract{We present the Bayesian method for evaluating the evidence for a non-zero value of the leptonic mixing angle $\theta_{13}$ and CP-violation in neutrino oscillation experiments. This is an application of the well-established method of Bayesian model selection, of which we give a concise and pedagogical overview. When comparing the hypothesis $\theta_{13} = 0$ with hypotheses where $\theta_{13} >0$ using global data but excluding the recent reactor measurements, we obtain only a weak preference for a non-zero $\theta_{13}$, even though the significance is over $3\sigma$. We then add the reactor measurements one by one and show how the evidence for $\theta_{13} > 0$ quickly increases. When including the \DC, \DB, and \RENO~data, the evidence becomes overwhelming with a posterior probability of the hypothesis $\theta_{13} = 0$ below $10^{-11}$. Owing to the small amount of information on the CP-phase $\delta$, very similar evidences are obtained for the CP-conserving and CP-violating hypotheses. Hence, there is, not unexpectedly, neither evidence for nor against leptonic CP-violation. However, when future experiments aiming to search for CP-violation have started taking data, this question will be of great importance and the method described here can be used as an important complement to standard analyses. 
}
\begin{document}
\maketitle

\section{\label{sec:Intro}Introduction}
The phenomenon of neutrino oscillations has been established in a
series of experiments using neutrinos from a wide range of sources, such as the Earth's atmosphere, the Sun, nuclear
reactors, and man-made accelerators \cite{Schwetz:2011qt,Schwetz:2011zk,Fogli:2011qn}. Neutrinos can only oscillate if they are massive, and there are many important questions raised related to neutrino masses, which are currently being investigated. 
These include the Dirac or Majorana nature of neutrino masses, and the question of whether total lepton-number is violated, information on which can be obtained through experiments searching for neutrinoless double beta decay \cite{KlapdorKleingrothaus:2000sn,Tretyak:2011zz,Andreotti:2010vj}. In addition, there is the quest for the determination of the absolute values of the neutrino masses, which is most directly performed using single beta decay experiments \cite{Kraus:2004zw,Troitsk:1999tp}, although cosmological observations can also provide information \cite{Komatsu:2010fb}. 

However, there are still many questions related to oscillation experiments which remain to be answered. The oscillations observed until recently correspond to the dominant effective two-flavor oscillation modes, driven by two mass-squared differences and two relatively large mixing angles, while the purpose of most current and future experiments is mainly to search for sub-leading effects. This includes the determination of the third leptonic mixing angle $\theta_{13}$ and of the neutrino mass ordering, which can either be normal or inverted. Also, the existence of CP-violation in the lepton sector is a very important question.

In fact, CP-violation in neutrino oscillations, which is a genuine three-flavour effect, is only possible for a
non-zero value of $\theta_{13}$, and any realistic possibility to determine the type of the neutrino mass
ordering relies on $\theta_{13}$ not being too small \cite{Mezzetto:2010zi}. Therefore, the results
on $\theta_{13}$ from the present generation of experiments will be of crucial importance for the feasibility of
future experiments aiming to determine the neutrino mass ordering or search for leptonic CP-violation.

For a long time, the neutrino oscillation data was largely consistent with a zero value of $\theta_{13}$.
However, a while ago, there was a series of experiments indicating a preference for a non-zero value of $\theta_{13}$ from the \MINOS~\cite{MINOS:2011}, \T2K~\cite{T2K:2011}, and \DC~\cite{DC:2011} collaborations. These measurements were recently followed by much more precise ones from \DB~\cite{DB:2012} and \RENO~\cite{RENO:2012}.

Due to the degeneracies between oscillation parameters in individual neutrino oscillation experiments, there exists a long history of global fits of oscillation data, with some of the most recent ones being Refs.~\cite{Schwetz:2011qt,Schwetz:2011zk,Fogli:2011qn}. These analyses apply standard frequentist methods in order to estimate the oscillation parameters and to determine the significance of $\theta_{13} > 0$ (see also Refs.~\cite{Escamilla:2008vq,Roa:2009wp}), but do not include any of the recent reactor data of the \DC, \DB, and \RENO~collaborations. Bayesian estimation of neutrino oscillation parameters using subsets of neutrino oscillation data has been performed in Refs.~\cite{Garzelli:2001zu,Ge:2008sj}. Finally, after the completion of this work, an update of Refs.~\cite{Schwetz:2011qt,Schwetz:2011zk} was made public, including the recent reactor data \cite{Tortola:2012a}.

In this work, we describe how to correctly determine the evidence for a non-zero value of the leptonic mixing angle $\theta_{13}$ and a non-trivial value of the CP-violating Dirac phase $\delta$ using Bayesian inference.
The well-established method of analyzing these types of questions is based on \emph{Bayesian model selection}, which we will describe in some detail. More information on model selection can be found in the textbooks and review articles in Refs.~\cite{Hobson:2010book,Sivia:1996,Trotta:2005ar,Trotta:2008qt,Kass:1995}. Applications within the fields of cosmology and astrophysics can be found in Refs.~\cite{Trotta:2008qt,Trotta:2005ar,Martin:2010hh,Arina:2011zh,Feroz:2010jc,Hobson:2010book}, while 
applications in particle physics are given in Refs.~\cite{Feroz:2008wr,AbdusSalam:2009tr,CLAS:2007,Cousins:2008,Ireland:2008}.
We will compare the hypothesis $\theta_{13} = 0$ with the hypotheses where $\theta_{13}$ can take on any value within its physical range, and in this way obtain the preference of a non-zero value of $\theta_{13}$. After assigning prior probabilities on the different hypotheses, we obtain posterior probabilities of the hypotheses that $\theta_{13}>0$ and $\theta_{13}=0$, respectively.
We evaluate how recent data has increased step by step the evidence so that it now overwhelmingly prefers a non-zero $\theta_{13}$ (which is not very surprising, given the recent reactor measurements). Furthermore, we automatically obtain the method of evaluating the Bayesian evidence for a non-trivial value of the phase $\delta$. Since current data is not very sensitive to the value of $\delta$, we find, not surprisingly, no evidence neither for nor against CP-violation in neutrino oscillations.
However, in the future when proposed experiments aiming to search for CP-violation (see, for example, Refs.~\cite{Bandyopadhyay:2007kx,Akiri:2011dv,Longhin:2011hn,Ayres:2004js}) have started giving data, this question will be of great importance, and the method described here can be used to assess to which degree that data favors or disfavors CP-violation in neutrino oscillations.

This work is organized as follows. In \secref{sec:BayesInf}, we review the principles of Bayesian inference, including a self-contained treatment of Bayesian model selection.
\secref{sec:MS_T13CPV} describes our method of performing model selection for determining the evidence favoring a non-zero $\theta_{13}$ and leptonic CP-violation, and includes model definitions, a thorough discussion on the choice of priors, and the approximation of the likelihood we need to make. The numerical results are presented, first using the global data analyzed in \refcite{Schwetz:2011zk}, and then also including the recent \DC, \DB, and \RENO~data. Finally, a brief summary and our conclusions are given in \secref{sec:summary}.

\section{Bayesian inference}\label{sec:BayesInf}
In the Bayesian interpretation, probability is associated with degree of belief.
This is in contrast to the frequentist interpretation, in which probability is defined as the limit of the relative frequency of an event in a large number of repeated trials.

Bayesian inference is a framework for updating prior belief or knowledge based on new information or data.  A common problem in data analysis is to use the data to make inferences about parameters of a given model, and at a higher level, to decide which of two or more competing models is preferred by the data.
Bayesian inference answers the question how probable a given value of a parameter, or a whole model, is, given the observed data.
If one is interested in probabilities of parameters within models, as well as the models themselves, one is forced to adopt the Bayesian approach. Generally, the probability $\Pr(A | B)$ represents the degree of belief regarding the truth of $A$, given $B$. The order of the conditioning can be reversed using \emph{Bayes' theorem},
\begin{equation} \Pr(A|B) =
\frac{\Pr(B|A)\Pr(A)}{\Pr(B)}.
\end{equation}

Often, one is interested to infer values of parameters of a model from a set of observations or data.
Given a model or hypothesis $H$ with a set of $N$ parameters $\mathbf{\Theta} = \lbrac \Theta_i \rbrac_{i=1}^N$, and a set of data $\mathbf{D}=\lbrac D_i \rbrac_{i=1}^M$, Bayes' theorem implies\footnote{All probabilities are also implicitly assumed to be conditioned on all the relevant background information $I$, \ie, $\Pr(X)$ is written instead of $\Pr(X|I)$.} 
\begin{equation} \label{eq:bayes} \Pr( \mathbf{ \Theta} | \mathbf{D},H) = \frac{\Pr(\mathbf{D}
|\mathbf{\Theta},H)\Pr(\mathbf{\Theta}|H)}
{\Pr(\mathbf{D}|H)}, 
\end{equation}
where $\Pr(\mathbf{\Theta}|\mathbf{D}, H)$ is the posterior probability (density) of the parameters $\mbTh$, given the model and the data, and $\pi(\mathbf{\Theta}) \equiv \Pr(\mathbf{\Theta}|H)  $ is the prior probability (density). The likelihood function $\mathcal{L}(\mathbf{\Theta}) \equiv \Pr(\mathbf{D}|\mathbf{\Theta}, H) $ is the probability (density) of the data $\mathbf{D}$, assuming parameter values $\mbTh$, while $\Pr(\mathbf{D}|H)$ is the \emph{Bayesian evidence} (or model likelihood), which is given by
\bea
\mathcal{Z} \equiv \Pr(\mathbf{D}|H) &=&  \int \Pr(\mathbf{D},\mbTh|H)\df^N\mathbf{\Theta} = \int \Pr(\mathbf{D}|\mbTh, H) \Pr(\mathbf{\Theta}|H)\df^N\mathbf{\Theta} \notag \\
&=& \int{\mathcal{L}(\mathbf{\Theta})\pi(\mathbf{\Theta})}\df^N\mathbf{\Theta},
\label{eq:Z}
\eea
and is simply the factor required to normalize the posterior in \eref{eq:bayes}.
Since the evidence does not depend on the values of the parameters $\mbTh$, it is usually ignored in parameter estimation problems and the parameter values and uncertainties are obtained using the unnormalized posterior. However, the evidence plays a central role in model selection, as will be described in \secref{sec:ModSelect}.

One important thing to keep in mind is that prior and posterior probability densities, in order to keep the total probability invariant, transform under a change of variables $\mbTh \ra \mbOm = \mbOm(\mbTh)$ by multiplication by the Jacobian determinant, \ie, as
\be \label{eq:vartrans} \Pr (\mbOm) = \Pr\left(\mbTh\right) \left| \frac{\partial\mbTh}{\partial\mbOm}\right|.\ee
Hence, a prior uniform in one parameter will not be so in a nonlinear function of it, and thus the specification of a prior is essentially equivalent to the specification of a variable in which the prior is uniform.

The probability density of any subset $\eta$ of the parameters $\mathbf{\Theta}=(\eta, \rho)$ is obtained by integrating over the other parameters $\rho$, fully taking into account their uncertainty, as
\be\Pr(\eta | X) = \int \Pr(\eta,\rho|X) \df \rho, \ee
for any $X$. This makes it possible to eliminate nuisance parameters in a fully consistent way by including them in the parameter space and then performing the above integral over the posterior distribution. In addition, the probability density of any (unique) function of the parameters  $K=F(\mbTh)$ is obtained as
\be\Pr(K|X)= \int \Pr(K |\mathbf{\Theta}, X) \Pr( \mathbf{\Theta} |X) \df^N \mathbf{\Theta} = 
\int \delta(K - F(\mathbf{\Theta})) \Pr( \mathbf{\Theta} |X) \df^N \mathbf{\Theta}.\ee
Although this might look like a daunting integral, if one has access to samples from $\Pr( \mathbf{\Theta} |X)$, one can easily find the total probability in an interval of $K$ by simply binning the samples. 

The main result of Bayesian parameter inference is the posterior and its marginalized versions (usually in one or two dimensions).
However, it is also common to give point estimates such as the posterior mean or median, as well as \emph{credible intervals (regions)}, which are defined as intervals (regions) containing a certain amount of posterior probability. Note that these regions are not unique without further restrictions, just as for classical confidence intervals. 

Examples of Bayesian analyses of particle physics models are global fits of supersymmetric extensions of the standard model \cite{Feroz:2008wr,Roszkowski:2009sm,Trotta:2008bp,Bertone:2011nj,AbdusSalam:2009tr,AbdusSalam:2009qd} and of the minimal universal extra dimensions scenario \cite{Bertone:2010ww}.

\subsection{Model selection \label{sec:ModSelect}}
The discussion in the previous section was concerned with the Bayesian method of inferring the values of parameters, \emph{assuming a certain hypothesis $H$ to be true}. However, another arguably much more important question is that of which hypothesis is the best description of the data in the first place. It is the evidence in \eref{eq:Z} which can be used to distinguish between a set of hypotheses $\lbrac H_i \rbrac_{i=1}^r$. This is because Bayes' theorem also implies 
\begin{equation}\label{eq:Bayes_model} \Pr(H_i|\mathbf{D}) =
\frac{\Pr(\mathbf{D}|H_i)\Pr(H_i)}
{\Pr(\mathbf{D})},
\end{equation}
and so gives the posterior ratio of probabilities of two hypotheses as
\begin{equation}\label{eq:post_ratio} \frac{ \Pr(H_i|\mathbf{D})}{\Pr(H_j|\mathbf{D})} =
\frac{\Pr(\mathbf{D}|H_i)}{\Pr(\mathbf{D}|H_j)} \frac{\Pr(H_i)}{\Pr(H_j)} = \frac{\mathcal{Z}_i}{\mathcal{Z}_j} \frac{\Pr(H_i)}{\Pr(H_j)},
\end{equation}
where $\Pr(H_i)/\Pr(H_j)$ is the prior probability ratio of the two models. The ratio of evidences, $B_{ij} = \ev_i/\ev_j$ is often called the \emph{Bayes factor}.
From \eref{eq:Z}, one observes that the evidence is the average of the likelihood over the prior, and hence this method automatically implements a form of \emph{Occam's razor}, since in general a simpler theory with a smaller parameter space will have a larger evidence than a more complicated one, unless the latter can fit the data substantially better.
More specifically, only the inclusion of parameters which are constrained by the data will lead to a smaller evidence, while the inclusion of parameters that are unconstrained will leave the evidence unaffected.
Since $\Pr(\mathbf{D}) = \sum_i \Pr(\mathbf{D} | H_i) \Pr(H_i)$, \eref{eq:Bayes_model} can easily be written in terms of the evidences and the prior probabilities as 
\be \label{eq:hyp_posterior} \Pr(H_j|\mathbf{D}) =  \frac{1}{1+\sum_{i \neq j}\frac{\mathcal{Z}_i}{\mathcal{Z}_j} \frac{\Pr(H_i)}{\Pr(H_j)}}. \ee

In the simplest case when two models being compared have no free parameters, the Bayes factor is simply the likelihood ratio. When the models have free parameters, it is still a likelihood ratio, but between the whole models, and this ratio is obtained by integrating over the parameter spaces of the models as in \eref{eq:Z}. Note that Bayesian model selection allows data to favor the simpler model. Also, it is possible to incorporate external information when comparing models.
In addition, note that the posterior distributions of the parameters do not depend on the overall scale of the likelihood, since the evidence scales accordingly. The same holds true for the posterior model probabilities, since the ratio of evidences in \eref{eq:hyp_posterior} is also independent of the likelihood normalization. The significance of the evidence is usually interpreted using \emph{Jeffrey's scale} in \tabref{tab:Jeffreys}, as used in, for example, Refs.~\cite{Trotta:2008qt,Trotta:2005ar,Hobson:2010book,Feroz:2008wr,AbdusSalam:2009tr}.

\begin{table}
\begin{center}
\begin{tabular}{|c|c|c|c|}
\hline
$\log(\ev_1/\ev_0)$ & $\ev_1/\ev_0$ & $\Pr(H_1 | \mathbf{D})$ & Interpretation \\ 
\hline
$<1.0$ & $\lesssim 3:1$ & $\lesssim 0.75$ & Inconclusive \\
$1.0$ & $\simeq 3:1$ &  $\simeq 0.75$ & Weak evidence \\
$2.5$ & $\simeq 12:1$ & $\simeq 0.92$ & Moderate evidence \\
$5.0$ & $\simeq 150:1$ & $ \simeq 0.993$ & Strong evidence \\ \hline
\end{tabular}
\end{center}
\caption{Jeffrey's scale often used for the interpretation of Bayes factors and model probabilities. The posterior model probabilities are calculated by assuming only two competing hypotheses and equal prior probabilities.}
\label{tab:Jeffreys}
\end{table}

Generally, as more data become available, for any (not completely unreasonable) prior distribution, the posterior becomes
practically independent of the prior, and determined solely by the likelihood. However, the dependence of the evidence on the prior always remains, although the Bayes factor will generally favor the correct model once \qu{enough} data has been obtained \cite{Trotta:2005ar}.

It can be interesting to compare this method with the completely different approach used in frequentist inference in the special case of nested models.
In the latter case, the significance of a new effect is usually evaluated using hypothesis tests \cite{Cowan:1998ji}. The null hypothesis $H_0$ of no effect is expressed as $\eta = \eta_0$ and is tested against the alternative hypothesis $\eta \neq \eta_0$. A significance level $\alpha$ is chosen and a test statistic with known probability distribution under $H_0$ is constructed. Using the observed data, an \qu{observed} value of the test statistic and a p-value $p$ is calculated. The null hypothesis is rejected if $p < \alpha$.\footnote{The p-value is not the significance of the test, but it is the highest significance at which $H_0$ could be rejected. This distinction will be implicitly assumed for the rest of this work.} Note that the p-value is not directly related to the probability of the null hypothesis being true.

A very commonly used statistic is based on the \emph{profile likelihood ratio}. Define
\be \label{eq:Qsquare} Q^2(\mathbf{D},\eta_0) \equiv - 2 \log  \frac{\sup_\rho\mathcal{L}(\eta_0,\rho)}{\sup_{\eta,\rho}\lhood(\eta,\rho)} =- 2 \log \frac{\lhood(\eta_0,\hat{\hat{\rho}}(\eta_0))}{\lhood(\hat{\eta},\hat{\rho})},  \ee
where \qu{$\sup$} denotes the supremum, a single hat denotes the parameters which maximize the likelihood, and a double hat indicates the conditional maximum for fixed $\eta_0$. Since in this case the dependence on the data $\mathbf{D}$ is important, it is explicitly shown as an argument. Under the assumption that $H_0$ is true, in the large sample limit and under some additional conditions, 
 $Q^2$ (often denoted by $\Delta \chi^2$) has a $\chi^2$-distribution with number of degrees of freedom equal to the dimensionality of $\eta$. This result is known as \emph{Wilks' theorem} \cite{Wilks:1938}. Frequentist \emph{confidence intervals} at confidence level $1-\alpha$ can then be constructed by performing the hypothesis test for all values of the parameter and then including all parameter values that are not rejected at a significance $\alpha$.  
This close relationship between between interval construction and hypothesis testing does not, however, have any analogy in Bayesian inference, \ie, there is no direct relation between parameter estimation and credible interval construction on the one hand and model selection on the other hand.
Note that a more complicated model will always be able to fit the data at least as good as the simpler model, and hence, unlike in Bayesian model selection, one can never obtain evidence \emph{in favor} of the null hypothesis. 

Sometimes it may happen that a conclusion based on a hypothesis test seems to contradict that obtained using model selection. This can, for example, happen if the estimate of $\eta$ is found far from the value of the null hypothesis and a large value of $Q^2$ is observed, and so the data is unlikely under $H_0$. However, the data could be even more unlikely under the alternative as in \eref{eq:Z}, and hence the Bayes factor can even favor $H_0$. Also, the prior probability of the alternative could be very low.
Thus, taking the Bayesian view also means that the significance, or the \qu{number of $\sigma$}, of a result is in general not a good indicator of the importance or the evidence of a new effect, a result that is known as \qu{Lindley's paradox}. For further details, see for example Appendix~A of \refcite{Trotta:2005ar}. As we will see, this situation does, to some extent, apply to the case of the third leptonic mixing angle $\theta_{13}$ when the recent reactor measurements are not taken into account.

We want to emphasize that the posterior \emph{probability density} at the special value $\eta = \eta_0$ of the alternative model says nothing about the probability that $\eta = \eta_0$, and neither does the relative posterior densities at $\eta = \eta_0$ and the maximum of the posterior. The reason is that, if one uses a continuous prior probability density for $\eta$, then the probability that $\eta = \eta_0$ is zero for any value of $\eta_0$. However, for nested models and under some additional assumptions, the ratio of the prior to the posterior at $\eta = \eta_0$ does indicate this probability, since it is in fact the Bayes factor \cite{Trotta:2005ar}.

Finally, we mention that the Bayesian analysis can be taken even one step further by performing \emph{model averaging}. Using this, the posterior for any set of parameters can be calculated, while taking into account the uncertainty regarding which model is the correct one. Once again, all one has to do is to use the laws of probability theory to obtain \cite{Hobson:2010book, Trotta:2008qt}
\be  \Pr(\mathbf{\Theta} |X) = \sum_{i=1}^r \Pr(\mathbf{\Theta} |H_i, X) \Pr(H_i |X), \ee
\ie, the probability distribution is given by the average of the individual distributions over the space of models, with weights equal to the  model probabilities.

\section{Bayesian model selection for $\theta_{13}$ and $\delta$}\label{sec:MS_T13CPV}
\subsection{Model definitions}
As the general framework, we simply take the Standard Model and augment it with a neutrino mass matrix. The nature of the mass matrix (Dirac or Majorana) is not important here, but we observe that the weakly interacting neutrino fields $\nu_{\alpha}$ ($\alpha = e, \mu, \tau$) are superpositions of mass eigenstate fields $\nu_i$ ($i = 1,2,3$) with masses $m_i$. The neutrino masses $m_i$ can either have normal ($m_1<m_2<m_3$) or inverted ($m_3<m_1<m_2$) ordering.
In a basis where the charged lepton mass matrix is
diagonal we have
\be \nu_\alpha = \sum_{i=1}^3 U_{\alpha i} \nu_i,\ee
where $U$ is the \emph{leptonic mixing matrix}, also referred to as the \emph{Pontecorvo-Maki-Nakagawa-Sakata} (PMNS) matrix, usually parametrized as
\begin{equation} \label{eq:U_PMNS}
\begin{split}
U &= \left( \begin{array}{ccc}
	1    &      0      &    0       \\
	0    &  c_{23} & s_{23}  \\
	0    & -s_{23} & c_{23}
\end{array} \right)	\left( \begin{array}{ccc}
	 c_{13}	&	0	&	s_{13}e^{-i\delta}	\\
	 0		&	1	&	0				\\
	-s_{13}e^{i\delta}  & 0 &	c_{13}
\end{array} \right)		\left( \begin{array}{ccc}
	 c_{12}	& s_{12}	& 0	\\
	-s_{12}	& c_{12}	& 0	\\
	  0		&  0		& 1
\end{array} \right)	
\mathrm{diag}\begin{footnotesize}\left(e^{i\rho},e^{i\sigma},1\right)\end{footnotesize} \\
& =  \left( \begin{matrix}c_{12} c_{13} & s_{12} c_{13} &
s_{13} e^{-{\rm i}\delta} \cr -s_{12} c_{23}-c_{12} s_{23} s_{13}
e^{{\rm i} \delta} & c_{12} c_{23}-s_{12} s_{23} s_{13} e^{{\rm i}
\delta} & s_{23} c_{13} \cr s_{12} s_{23}-c_{12} c_{23} s_{13}
e^{{\rm i} \delta} & -c_{12} s_{23}-s_{12} c_{23} s_{13} e^{{\rm i}
\delta} & c_{23} c_{13}\end{matrix} \right) \left(
\begin{matrix} e^{{\rm i}\rho} & 0 & 0 \cr 0 & e^{{\rm i}\sigma} & 0 \cr 0 & 0 &
1 \end{matrix} \right) \\
\end{split} 
\end{equation}
where $c_{ij} = \cos \theta_{ij}$ and $s_{ij} = \sin \theta_{ij}$, $\theta_{12}, \theta_{23},$ and $\theta_{13}$ are the lepton mixing angles, $\delta$ is the CP-violating Dirac phase, and $\sigma$ and $\rho$ are CP-violating Majorana phases, which are only relevant in the case of Majorana neutrinos. The physical ranges for the mixing angles are the intervals $[0,\pi/2]$ \cite{Lundell:1999kn}, while $\delta$ can be restricted to $[-\pi, \pi]$, and the Majorana phases to $[0,\pi]$.
With $\Delta m_{21}^2 = m_2^2 - m_1^2$ and $\Delta m_{31}^2 = m_3^2 - m_1^2$, neutrino oscillation experiments are sensitive to the set of parameters $ \mbTh_{\text{osc}} = (\Delta m_{21}^2,\Delta m_{31}^2, \theta_{12}, \theta_{23}, \theta_{13}, \delta) $,
meaning that these are the parameters that the oscillation probabilities depend on. Defining
\bea \psi &=& \left(  \Delta m_{21}^2,\Delta m_{31}^2, \theta_{12}, \theta_{23} \right), \eea
we distinguish the two CP-conserving cases $\delta=0$ and $\delta= \pi$, and define the following hypotheses 
\bea H_0: \mathbf{\Theta}_0 &=&  \psi, \; \theta_{13} = 0 \quad (\delta \text{ irrelevant}), \\
H_1: \mathbf{\Theta}_1 &=& (\psi, \theta_{13}), \; \theta_{13} \in [0, \pi/2], \; \delta=0,  \\
H_2: \mathbf{\Theta}_2 &=& (\psi, \theta_{13}), \; \theta_{13} \in [0, \pi/2], \;  \delta = \pi  \quad (\text{ or } -\pi), \\
H_3: \mathbf{\Theta}_3 &=& (\psi, \theta_{13}, \delta ), \;  \theta_{13} \in [0, \pi/2], \;  \delta \in [-\pi,\pi].
\eea
These hypotheses can then be compared by computing the evidences in \eref{eq:Z}. For this one requires the likelihood function and priors on the oscillation parameters.  These will be described in Secs.~\ref{sec:ev_calculation} and \ref{sec:priors}, respectively.

The neutrino mass ordering will be treated as fixed, and the very small differences obtained by assuming the different orderings to be true will be evaluated in \secref{sec:results}. However, once more data from future experiments searching for CP-violation become available, the constraints on the phase $\delta$ is expected to depend very much on the assumed mass ordering \cite{Huber:2009cw}. In this case, it might be preferable to define separate models for each mass ordering, and perform model selection on all eight models simultaneously.

\subsection{Choice of priors}\label{sec:priors}
In any Bayesian analysis, one needs to specify prior probability distributions on the parameters in a given model, and if more than one model is considered, probabilities on the space of models. In general, the prior should reflect ones prior knowledge, given the relevant background information. However, some difficulties can appear if there is very little background information, or if it is difficult to translate this information into mathematically precise statements. 

First, we consider the question of how to assign prior model probabilities. A reasonable choice could be to use prior probabilities $\Pr(H_0) = 1/3$, $\Pr(H_1) = \Pr(H_2) = 1/6 $, and $\Pr(H_3) = 1/3$. Then, however, the a priori probability of $\theta_{13} = 0$ would only be $1/3$ rather than $1/2$, which could be considered more natural. This will, as always, only have a non-negligible effect on the posterior probabilities (as it should) when the data is not very informative. Even in this case the effect will be small, and actually much smaller than the effect coming from the variations in the evidences when different priors of $\theta_{13}$ are chosen.

Then, there is the task of assigning prior probability densities on the continuous parameters of the models.
To deal with this, a wide variety of methods and rules have been developed which one can use to obtain priors in cases when there is little prior information (see, for example, Refs.~\cite{Kass:1996,Berger:1985}). These methods can serve as a guide to what shapes a reasonable distribution might have. However, we take the common  point of view that, in general, these priors should not be accepted too blindly. In the end, the priors should be proper, \ie, normalized to unity, and be a reasonable reflection of one's degree of belief. By considering the variation of the posterior inference when a set of different such priors is considered, one can check the robustness of the obtained results.

We assume a prior on the form
\be \label{eq:priorsep} \pi(\psi,\theta_{13}, \delta) = \pi(\psi)\pi(\theta_{13}) \pi(\delta)\ee
on the oscillation parameters. Besides being quite natural, this form of the prior can be obtained by demanding that the prior must not depend on which basis the neutrino mass matrix is defined in. The resulting so-called \emph{Haar measures} on the leptonic mixing matrix have been discussed in Refs.~\cite{Haba:2000be,Espinosa:2003qz},\footnote{However, note the arising complications due to the unphysical phases discussed in \refcite{Espinosa:2003qz}.} which lead to unique and separable priors on the mixing angles and phases. The prior on $\psi$ is taken to be the same for all hypotheses.

\begin{figure}[t]
\centering
\includegraphics[width=0.8\textwidth]{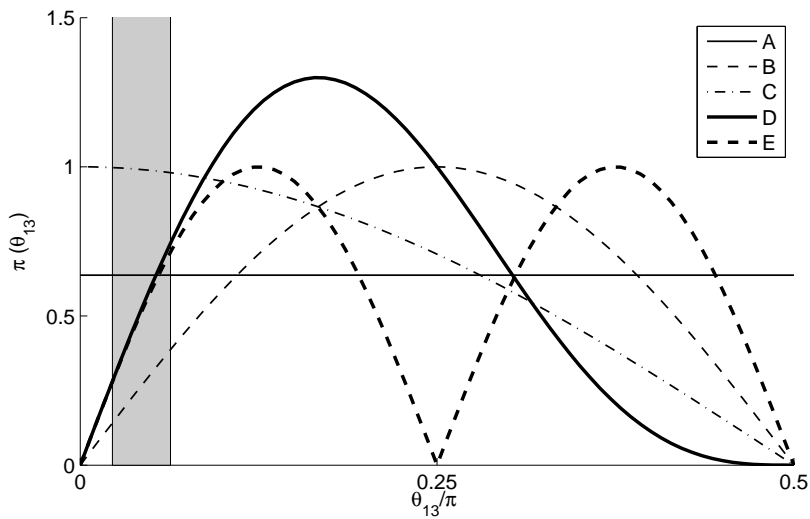}
\caption{The different correctly normalized priors on $\theta_{13}$. The grey band marks the approximate region from which the main contribution to the evidence originates.}\label{fig:priors}
\end{figure}

Our form of the evidences, derived in \secref{sec:ev_calculation}, will be independent of the prior on $\psi$, and hence from now on we only consider the priors on $\theta_{13}$ and $\delta$. In principle, one could simply make up a list of different reasonable priors, or equivalently variables in which one chooses a uniform prior, to investigate.  The different choices of priors on $\theta_{13}$ considered in this work are those derived from the Haar measures on SO$(2)$, U$(2)$, SO$(3)$, and U$(3)$, respectively.\footnote{That is, corresponding to real two-flavor, complex two-flavor, real three-flavor, and complex three-flavor mixing, respectively.} Also, since the experiments most sensitive to $\theta_{13}$, \ie, those involving accelerator and reactor neutrinos, are mainly sensitive to $\sin^2(2\theta_{13})$, one can choose a prior uniform in $\sin^2(2\theta_{13})$, in which case \eref{eq:vartrans} is used to extend the prior to the whole physical range $\theta_{13} \in [0,\pi/2]$. 
The different priors can be summarized as
\bea
\mathcal{A}:& & \pi(\theta_{13}) = 2/\pi \notag \\
\mathcal{B}:& & \pi(s_{13}^2) = 1 \notag \\
\mathcal{C}:& & \pi(s_{13}) = 1 \\
\mathcal{D}:& & \pi(c_{13}^4) = 1 \notag \\
\mathcal{E}:& & \pi\left(\sin^2(2\theta_{13})\right) =  1 \notag,
\eea
where $\theta_{13} \in [0, \pi/2]$ and all other variables are in the interval $[0,1]$. The corresponding implied priors on $\theta_{13}$ according to \eref{eq:vartrans} are plotted in \figref{fig:priors}, where also the approximate region in which most of the contribution to the evidences originates from is marked with a grey band. It is important to note that we do not consider any of the studied priors to be \qu{better} than the others, but instead consider them all as reasonable, and then evaluate how much the posterior inferences depend on the choice of priors.

Note that, in the rough approximation of a box-shaped likelihood and a constant prior within this region, the evidence is proportional to that constant value of the prior. Hence, one can estimate how the evidence for models with non-zero $\theta_{13}$ will depend on the prior chosen. 
Most importantly, the prior $\mathcal{C}$ will give the largest evidence, approximately a factor of two to four larger than $\mathcal{B}$, which yields the smallest evidence. This will be checked numerically in \secref{sec:results}.

For the model $H_3$, one also needs to specify a prior on $\delta$. However, since the constraints on $\delta$ are so weak, not only will its inclusion in the first place effect the evidence very little, but also the form of its prior will be largely irrelevant. The Haar measure is uniform in $\delta$, and so we use $\pi(\delta) = 1/(2\pi)$. If, in the future, the data becomes more informative regarding the value of $\delta$, one would need to evaluate if that data indeed supports the existence of CP-violation in neutrino oscillations, and the general method presented in this work can be followed. In this case, one could also evaluate the sensitivity to the choice of the prior on $\delta$. 
Since it is $\sin{\delta}$ and $\cos{\delta}$ which appears in the leptonic mixing matrix in \eref{eq:U_PMNS}, and hence also enter into the oscillation probabilities and CP-asymmetries, suitable choices could be priors uniform in $\sin \delta$ and $\cos \delta$.

\subsection{Approximate Bayes factors}\label{sec:ev_calculation}
Ideally, one would like to use the full likelihood of all the oscillation parameters, as well as the parameterization of the systematic uncertainties.
However, since we do not have the machinery to do this, we are forced to make some simplifying assumptions.
For any individual oscillation experiment, the likelihood as a function of all the oscillation parameters is in general highly degenerate and/or multi-modal. Nevertheless, thanks to the large number of different types of experiments, imposing constraints on different combinations of oscillation parameters, the likelihood has become unimodal and, at least to some approximation, Gaussian. The correlations between the different oscillation parameters in the standard parameterization are small, and the best-fit values of the different oscillation parameters are largely independent of the values of the other parameters \cite{Schwetz:2011qt,Schwetz:2011zk,Fogli:2011qn}.
We thus neglect any dependence between $\psi$ as a whole and the pair $(\theta_{13},\delta)$ and approximate the likelihood as
\be \label{eq:Lapprox} \mathcal{L}(\psi,\theta_{13},\delta) \simeq \lhood^{\text{N}}(\psi) \lhood^{\text{CPV}}(\theta_{13},\delta). \ee
With this approximation, the dependence of $\psi$ factorizes out, and the two-dimensional marginalized posterior is simply
\be \Pr(\theta_{13},\delta | \mathbf{D}, H_3) \propto \lhood^{\text{CPV}}(\theta_{13},\delta)\pi(\theta_{13},\delta).\ee
Furthermore, from \eref{eq:Qsquare}, one easily obtains
\be Q^2(\theta_{13},\delta) = -2\log\left(\frac{\lhood^{\text{CPV}}(\theta_{13},\delta)}{\lhood^{\text{CPV}}(\hat{\theta}_{13},\hat{\delta})}\right),\ee
giving
\be \label{eq:likeCPV}\lhood^{\text{CPV}}(\theta_{13},\delta) = \lhood^{\text{CPV}}(\hat{\theta}_{13},\hat{\delta})\exp\left(-\frac{Q^2(\theta_{13},\delta)}{2}\right) \ee
since the conditional maximum with respect to $\psi$ is independent of $\theta_{13}$ and $\delta$ in this case. The quantity $\lhood^{\text{CPV}}(\hat{\theta}_{13},\hat{\delta})$ is an irrelevant factor which can be set to unity.
The likelihood $\lhood^\text{CPV}$ without recent reactor data (we denote this data set by GF) is evaluated using \eref{eq:likeCPV}, and a map of $Q^2$ in the $ \theta_{13} -\delta$ plane obtained from the authors of \refcite{Schwetz:2011zk}.\footnote{This corresponds to the \qu{recommended} analysis of \refcite{Schwetz:2011zk}, which includes the short-baseline reactor data.}

Since the numbers of events in the reactor experiments \DC~(DC), \DB~(DB), and \RENO~are quite large \cite{DC:2011,DB:2012,RENO:2012}, the likelihoods are taken as Gaussian functions in the mean number of events, or equivalently, in $\sin^2(2 \theta_{13})$, 
\bea \sin^2(2\theta_{13}) &=& 0.086 \pm 0.0508 \quad \text{ (DC)}, \\ 
\sin^2(2\theta_{13}) &=& 0.092 \pm 0.0177 \quad \text{ (DB)}, \\
\sin^2(2\theta_{13}) &=& 0.113  \pm 0.0230 \quad \text{ (RENO)},
\eea
where the systematic errors\footnote{We assume the systematic errors to be independent. Although this is not really correct, it should be a sufficiently good approximation for our purposes.} have been integrated over using Gaussian distributions, giving the resulting effective likelihoods as Gaussians with unchanged means and deviation parameters added in quadrature.

From \eref{eq:Z} and with the approximation in \eref{eq:Lapprox}, one finds
\bea \ev_0 &=& \int \lhood(\psi,0,0)\pi(\psi) \df \psi \notag \\
&\simeq& \int \lhood^\text{N}(\psi)\pi(\psi)\df \psi  \cdot \lhood^\text{CPV}(0,0), \\
\ev_3 &=& \int \lhood(\psi,\theta_{13}, \delta)\pi(\psi,\theta_{13}, \delta) \df \psi \df \theta_{13} \df \delta \notag \\
&\simeq& \int \lhood^\text{N}(\psi)\pi(\psi)\df \psi  \cdot  \int \lhood^\text{CPV}(\theta_{13}, \delta) \pi(\theta_{13}, \delta)  \df \theta_{13} \df \delta,
\eea
so that the ratio of the evidences is given by
\be \frac{\ev_3}{\ev_0} \simeq \frac{\int \lhood^\text{CPV}(\theta_{13}, \delta) \pi(\theta_{13}, \delta)  \df \theta_{13} \df \delta}{\lhood^\text{CPV}(0,0)}.\ee
In a similar way, one obtains
\bea \frac{\ev_1}{\ev_0} &\simeq& \frac{\int \lhood^\text{CPV}(\theta_{13}, 0) \pi(\theta_{13})  \df \theta_{13}}{\lhood^\text{CPV}(0,0)},\\
\frac{\ev_2}{\ev_0} &\simeq& \frac{\int \lhood^\text{CPV}(\theta_{13},\pi) \pi(\theta_{13})  \df \theta_{13}}{\lhood^\text{CPV}(0,0)}.
\eea
Although the approximation in \eref{eq:Lapprox} certainly introduces some error on the calculated evidences, it is expected to be much smaller than the uncertainty coming from the choice of prior distributions, which is often relatively large.

The required integrals can, due to the simple form of the likelihood $\lhood^\text{CPV}$ and the low dimensionality, be evaluated using standard integration routines. For more complicated evidence integrals, more specific algorithms such as \MN~\cite{Feroz:2007kg,Feroz:2008xx}, which we also use in this work, are required.

\subsection{Numerical results}\label{sec:results}
\begin{figure}[t]
\centering
\begin{tabular}[t]{lr}
\includegraphics[width=0.5\textwidth]{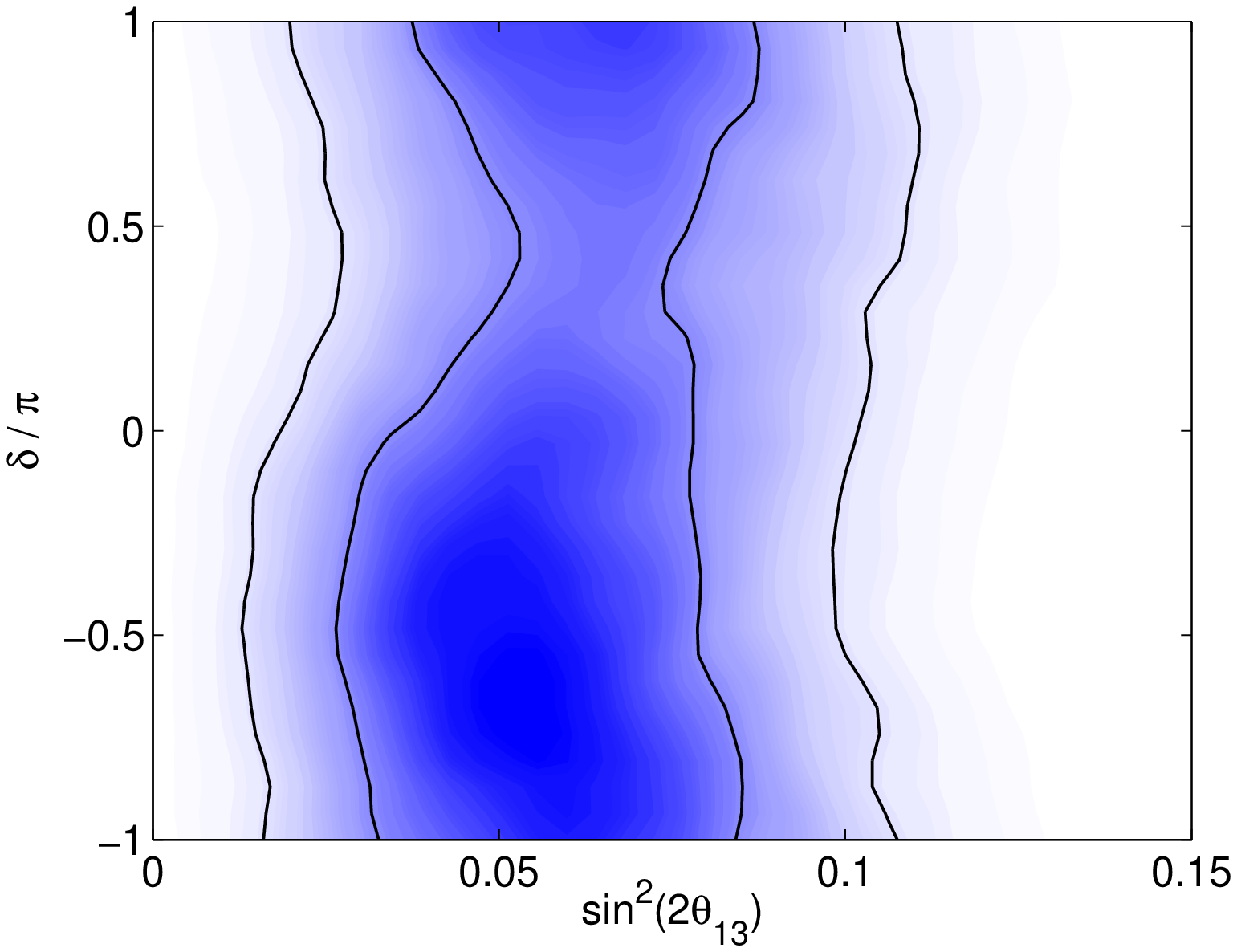}
\includegraphics[width=0.5\textwidth]{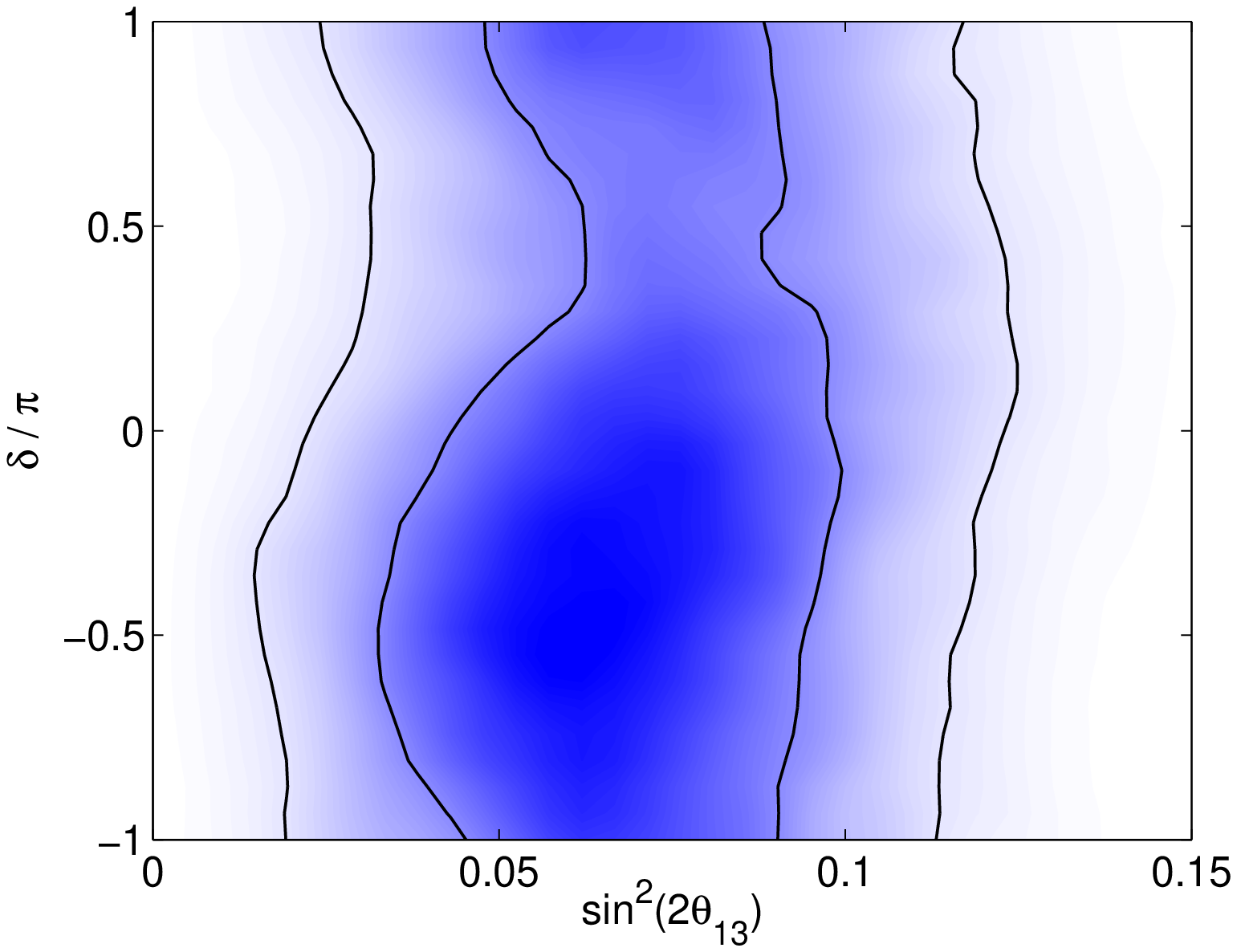}
\end{tabular}
\caption{Two-dimensional posterior distributions for $\sin^2(2 \theta_{13})$ and $\delta$, using uniform priors on these variables and the data analyzed in \refcite{Schwetz:2011zk}. Left (right) plots are for normal (inverted) neutrino mass ordering. The two black curves enclose $68~\%$ and $95~\%$ of the posterior probability, respectively.}\label{fig:2Dpost_GF}
\begin{tabular}[t]{lr}
\includegraphics[width=0.5\textwidth]{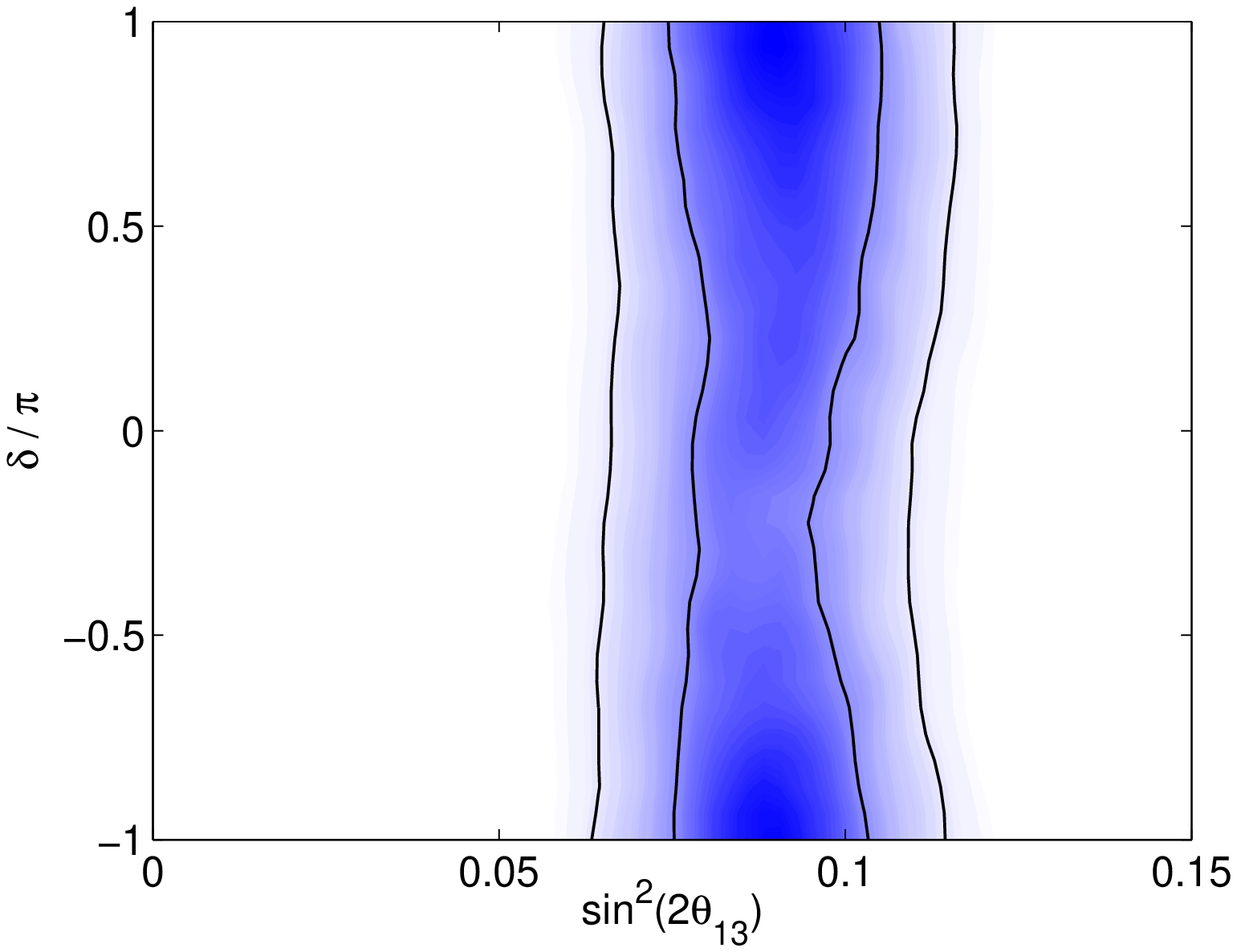}
\includegraphics[width=0.5\textwidth]{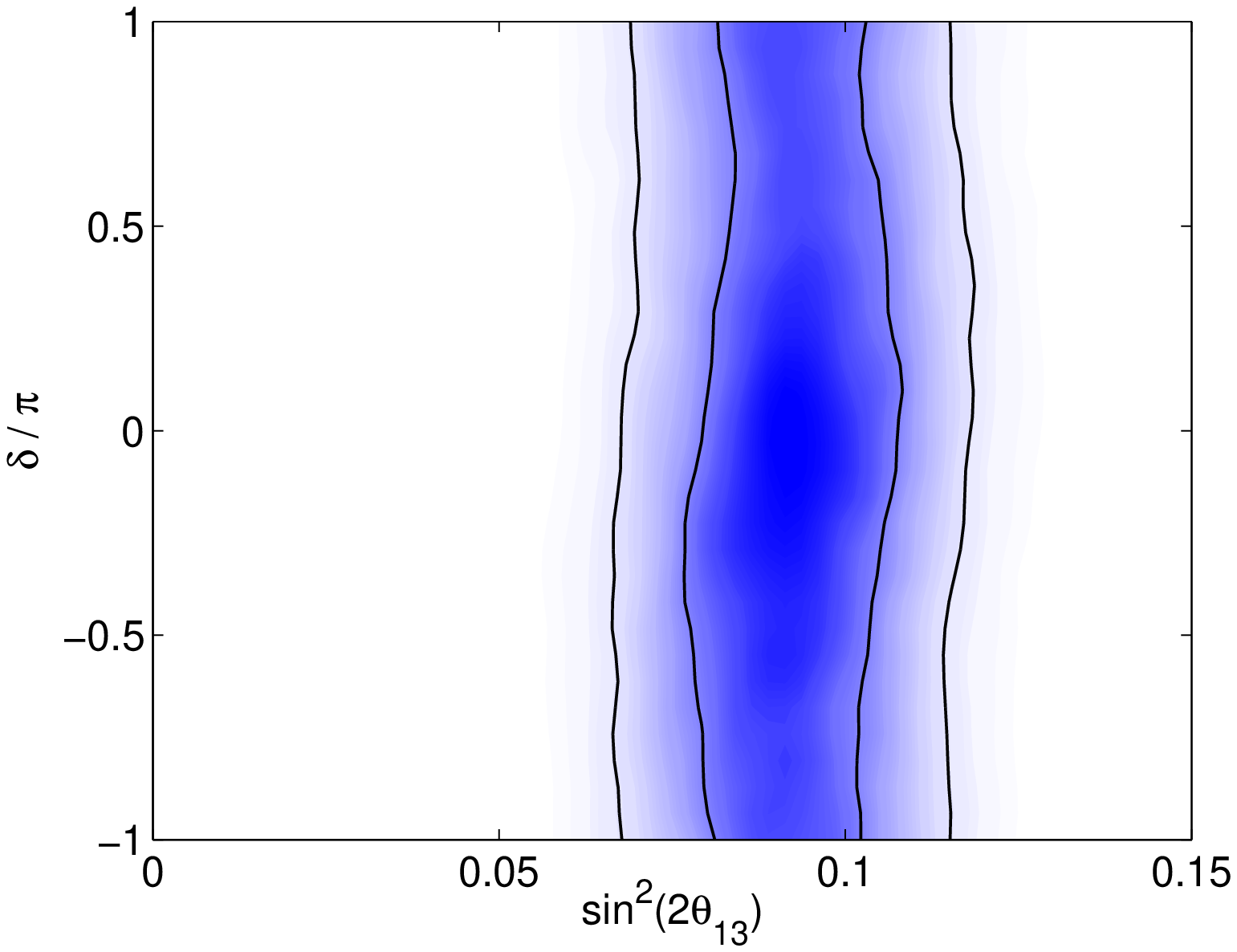}
\end{tabular}
\caption{Same as \figref{fig:2Dpost_GF}, but using all available data.}\label{fig:2Dpost_all}
\end{figure}

\begin{figure}[t]
\centering
\begin{tabular}[t]{lr}
\includegraphics[width=0.8\textwidth]{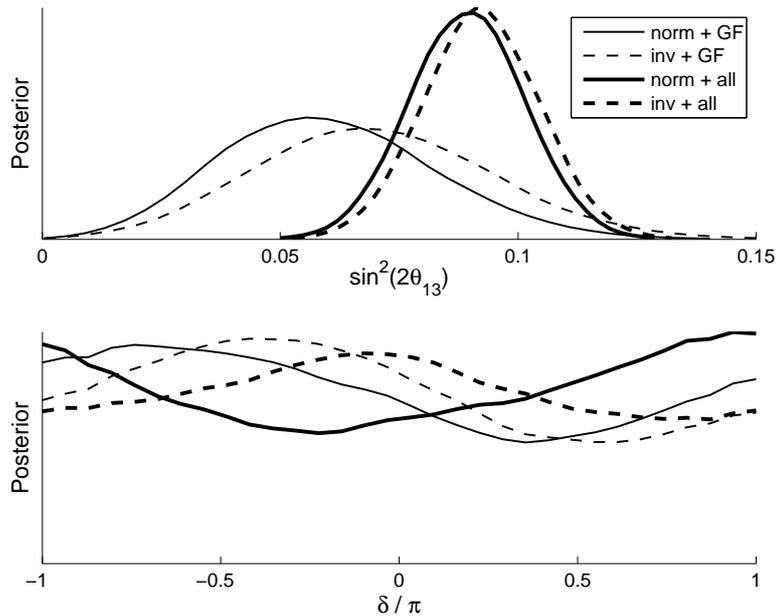}\\
\end{tabular}
\caption{One-dimensional posterior distributions of $\sin^2(2 \theta_{13})$ and $\delta$ for different combinations of mass orderings and data sets.}\label{fig:1Dpost}
\end{figure}

Our reconstructed two-dimensional posteriors using the data analyzed in \refcite{Schwetz:2011zk}, \emph{assuming $H_3$ to be true}, and using prior $\mathcal{E}$, are shown in \figref{fig:2Dpost_GF} for normal (left) and inverted (right) mass orderings. The $68~\%$ and $95~\%$ equal-posterior credible regions are marked by the black contours. Using different priors on $\theta_{13}$ results in very similar posteriors. In \figref{fig:2Dpost_all}, the posteriors including all the recent reactor data are shown. Besides the obvious observation that now $\sin^2(2\theta_{13})$ is much better constrained, one notices that the preferred values of $\delta$ have changed. Although the added reactor data is completely insensitive to the value of $\delta$, the degeneracy between $\delta$ and $\sin^2(2\theta_{13})$ in the other data results in changes of the preferred values. The maximum likelihood estimates are now $\hat{\delta} \simeq \pi$ and $\hat{\delta} \simeq 0$ for normal and inverted mass orderings, respectively. The one-dimensional marginalized posteriors are shown in \figref{fig:1Dpost}.
Our obtained maximum likelihood estimates and $68~\%$ confidence intervals of $\sin^2(2\theta_{13})$ when using all available data are $\sin^2(2\theta_{13}) = 0.090 \pm 0.012 $ and $\sin^2(2\theta_{13}) = 0.092  \pm 0.013$ for normal and inverted mass ordering, respectively.\footnote{The more careful analysis of \refcite{Tortola:2012a} do not use the short-baseline reactor data and finds somewhat larger preferred values of $\theta_{13}$.} Evaluating the posterior means and $68~\%$ credible intervals yields essentially the same values.

Note that the peaks of these posteriors, which are obtained \emph{assuming that $\theta_{13}$ is non-zero}, are quite far from $\theta_{13} = 0$, but in order to quantify the evidence of a non-zero $\theta_{13}$ one has to perform model selection. 
Due to the small amount of information regarding the value of $\delta$, the evidences for the models $H_1, H_2$, and $H_3$ are very similar, $|\log(\ev_{2,3}/\ev_1)| \lesssim 0.2$ without \DB~and \RENO~data, and $|\log(\ev_{2,3}/\ev_1)| \lesssim 0.5$ when including all data.\footnote{This does not depend on the prior on $\theta_{13}$.} The difference in the logarithms of the Bayes factors when assuming the different mass orderings to be true (around $0.2$ in log evidence) can be used as an estimate of the error on these logarithms induced by assuming the wrong mass ordering.

First, the analysis is done using only the data analyzed in \refcite{Schwetz:2011zk}. After that, we calculate how the evidence for $\theta_{13} > 0$ changes when the \DC, \DB, and \RENO~result are added, respectively. In \tabref{tab:evidences}, the evidences $\ev \simeq \ev_1 \simeq \ev_2 \simeq \ev_3$ compared to $\ev_0$ are shown. The ranges given are those obtained using the different priors on $\theta_{13}$, but also the smaller variations coming from assuming different mass orderings are included.\footnote{The numerical errors on all logarithms of Bayes factors in this work are about $0.05$ or smaller and can hence safely be neglected.} 
Furthermore, the posterior probability of $H_0$ is shown, calculated assuming that $\Pr(H_0) = 1/2$ and $\Pr(H_0) = 1/3$ (in the parentheses), respectively. Note that, due to the similarity of the evidences $\ev_1, \ev_2$, and $\ev_3$ of the other models, the posterior probability of $H_0$ is independent of the prior probabilities of these models for fixed $\Pr(H_0)$.
In the last column, the square root of the test statistic $Q^2$ is given for testing $H_0$ against $H_1$ (but with very small differences if testing against the other models), which would equal the \qu{number of $\sigma$} under the conditions of Wilks' theorem. 
 
First, we only use the data included in the global fit of Ref.~\cite{Schwetz:2011zk} (again denoted by GF). Although there is more than a $3 \sigma$ significance, there is in fact only weak evidence of $\theta_{13} >0$, with logarithms of the Bayes factors lying in the range $0.4 - 2.1$ for the different priors on $\theta_{13}$. The posterior probability of $H_0$, for this range of evidences, is in the range $0.11 - 0.39$ if $\Pr(H_0) = 1/2$ and in the range $0.06 - 0.24$ if $\Pr(H_0) = 1/3$.
One can then also include the variation of the posterior probability of $H_0$ coming from the variation of $\Pr(H_0)$ between $1/3$ and $1/2$ to obtain the range $0.06-0.39$ for the posterior of $H_0$, including all discussed uncertainties. 
As discussed in \secref{sec:priors}, the prior $\mathcal{B}$ yields the smallest evidence due to the fact that it puts very little prior probability into the high-likelihood region (see \figref{fig:priors}). 

When adding the \DC~measurements, the evidence for a non-zero $\theta_{13}$ increases. However, there is still no conclusive evidence. All logarithms of Bayes factors increase by about $1$ or slightly more for the different priors, yielding the range $1.7 - 3.1$ for $\log(\ev/\ev_0$) and posterior probabilities of $H_0$ (when including all uncertainties as in the previous paragraph) in the range $0.02-0.15$. The significance found is about $3.5 \sigma$.

Only the \DB~data by itself yields a significance of $5.2 \sigma$, and also the Bayesian evidence strongly prefers a non-zero $\theta_{13}$.  The logarithms of Bayes factors are now all larger than $9$, equivalent to posterior probabilities of $H_0$ smaller than $10^{-4}$. When the \DB~data is combined with the previous data, we find that $\log(\ev/\ev_0) > 14$, yielding posterior probabilities smaller than $7\cdot 10^{-7}$ for all prior combinations. The total significance is about $6.1 \sigma$.

Finally, when all the available data is included, we find that the significance is about $7.8 \sigma$, and the evidences $\log(\ev/\ev_0) > 25$, giving posterior probabilities of $H_0$ smaller than $10^{-11}$ for all combinations of priors.

\begin{table}[t]
\begin{center}
\begin{tabular}{|c|c|c|c|c|}
\hline
Data used & $\log(\ev/\ev_0)$ & $\ev/\ev_0$ & $\Pr(H_0 | \mathbf{D})$ & $\sqrt{Q^2}$ \\ 
\hline
\hline
GF  & $0.4-2.1$ & $1.6 - 7.9 $ & $0.11 (0.06)- 0.39 (0.24)$ & 3.1 \\
\hline
GF + DC & $1.7 - 3.1$ & $ 5 - 22$ & $0.04 (0.02) - 0.15 (0.08)$ & 3.5 \\
\hline
GF + DC + DB & $14.2 - 15.4$ & $ (1.5 - 5)\cdot 10^6$ & $< 7\cdot 10^{-7}$ & 6.1 \\
\hline
All & $25.3 - 26.5$ & $ (1-3) \cdot 10^{11} $ & $< 10^{-11}$ & 7.8 \\
\hline
\hline
Only DB & $9.1 - 10.3$ & $ (9 - 30)\cdot 10^3$ & $< 10^{-4}$ & 5.2 \\
\hline
\end{tabular}
\end{center}
\caption{Summary of Bayes factors and posterior probabilities of $H_0$ for $\Pr(H_0) = 1/2$ (and $\Pr(H_0) = 1/3$ within parenthesis). The ranges of $\ev \simeq \ev_1 \simeq \ev_2 \simeq \ev_3$ given are those obtained using the different priors on $\theta_{13}$, but also the smaller variation coming from the assumed mass ordering is included. The last column gives the square root of the observed $Q^2$ (only $H_0$ against $H_1$), which would equal the \qu{number of $\sigma$} under the conditions of Wilks' theorem.}\label{tab:evidences}
\end{table}

We observe that, for the first two cases in \tabref{tab:evidences}, the p-values corresponding to the observed values of $Q^2$ (which, again, are directly unrelated to the model probabilities) are rather small, while the posterior probabilities of $H_0$ are still significantly non-zero. This is the automatic Occam's razor effect of model selection at work: although the hypotheses with non-zero $\theta_{13}$ can accommodate significantly better fits, this is expected since these hypotheses have more freedom in their predictions, and this lack of predictivity is punished when using model selection. Although the maximum likelihoods are very large, the average ones are not, and these are what matters in model selection. Hence, the simpler model with $\theta_{13} = 0$, being more predictive, is not strongly disfavored. The situations are different for the three cases in the last three rows of \tabref{tab:evidences}. There, the likelihoods are so much larger for the alternative models that they are very strongly preferred, even though they are still less predictive and hence punished by the Occam's razor effect.

Since the posterior probability of $H_0$ is so incredibly small when including all data and in practice completely independent of any prior assumptions, we conclude that $\theta_{13}$ is non-zero with a probability practically equal to one. The next question is then if there is evidence of CP-violation in neutrino oscillations, \ie, a non-trivial value of the phase $\delta$. 
Since the three models $H_1, H_2,$ and $H_3$ all have $\theta_{13}$ as a free parameter, the resulting Bayes factors will to a very good approximation be independent of the prior on $\theta_{13}$. Using all available data, we obtain the Bayes factors for normal mass ordering
\be \log(\ev_2 / \ev_1) \simeq 0.5, \quad \log(\ev_3 / \ev_1) \simeq 0.2,\ee 
while for the inverted ordering, the result is
\be \log(\ev_2 / \ev_1) \simeq -0.2, \quad \log(\ev_3 / \ev_1) \simeq -0.3.\ee 
Hence, there is evidence neither for nor against CP-violation in neutrino oscillations. This was of course expected since the likelihood depends very weakly on $\delta$.
If one assigns prior probabilities as in \secref{sec:priors}, \ie, $\Pr(H_0) = 1/3, \Pr(H_1) = \Pr(H_2) = 1/6 $, and $\Pr(H_3) = 1/3$, the posterior probability of $H_3$ is very close to $0.5$ for both mass orderings ($|\Pr(H_3 | \mathbf{D})-0.5| \lesssim 0.05$), while the rest of the probability is shared between the $H_1$ and $H_2$ in slightly different proportions for the different mass orderings.

\section{Summary and conclusions}\label{sec:summary}
We have described how to determine the evidence of a non-zero value of the leptonic mixing angle $\theta_{13}$ and a non-trivial CP-violating Dirac phase from neutrino oscillation data, using Bayesian model selection. After having given a short, and hopefully clear, summary of the principles of Bayesian inference in general, and model selection in particular, we have applied it to the case of $\theta_{13}$ and $\delta$.

We have compared the hypothesis $\theta_{13} = 0$ with hypotheses where $\theta_{13}$ can take on any value within its physical range and, by calculating the Bayesian evidences, obtained the evidence favoring a non-zero $\theta_{13}$. After assigning prior probabilities to the different hypotheses, we have calculated their posterior probabilities. 
We have shown how recent data has step by step increased the evidence disfavoring $\theta_{13} = 0$, so that there is now overwhelming evidence of $\theta_{13}$ being non-zero.

Although the strong preference for a non-zero $\theta_{13}$ was expected, given the recent reactor data, this analysis also serves other purposes. Most importantly, we have described the Bayesian way of evaluating to what extent oscillation data supports the existence of CP-violation in neutrino oscillations through a non-trivial value of the phase $\delta$. 
Since there is not much sensitivity to the value of $\delta$ in current data, we find, not surprisingly, no evidence neither for nor against CP-violation.
However, in the future, when proposed experiments aiming to measure CP-violation have started taking data, this question will be of great importance.
The method described here can be used to assess to which degree that future data favors CP-violation in neutrino oscillations, and act as an important complement to standard $\chi^2$-analyses. In fact, it would be very interesting to evaluate how much proposed experiments would be able to provide (Bayesian) evidence of CP-violation (following, for example, Refs.~\cite{Trotta:2010ug,Trotta:2007hy,Watkinson:2011xp}).

Finally, we note that Bayesian model selection could also be applied to other problems related to neutrino oscillations. For example, it could be used to decide which mass ordering of the neutrinos is preferred. To do this one would essentially need the full likelihood of all the oscillation parameters and make sure that one uses the same normalization of the likelihood throughout. Since future constraints on the phase $\delta$ is expected to depend very much on the assumed mass ordering, the best approach might be to define separate models for each assumption on $\delta$ and for each mass ordering, and then to perform model selection on all models simultaneously. In addition, one could use model selection to investigate whether $\theta_{23}$ is maximal or not, or if oscillation data requires the introduction of sterile neutrinos.

\acknowledgments
The author would like to thank T.~Schwetz, M.~T{\'o}rtola, and J.~Valle for sharing their profile likelihood as a function of $\theta_{13}$ and $\delta$, and in addition M.~P.~Hobson, A.~Merle, and T.~Ohlsson for helpful comments on the manuscript.

\bibliographystyle{JHEP}
\bibliography{T13ModSelect}

\end{document}